\documentclass[aps,prb,twocolumn,superscriptaddress,groupedaddress]{revtex4}
\usepackage{graphicx}
\usepackage{fancyhdr}
\usepackage{amsmath}
\usepackage{xcolor}
\usepackage[colorlinks=true, allcolors=blue]{hyperref}

\begin{document}

\preprint{preprint}

\title{Resonant pinning spectroscopy with spin-vortex pairs}

\author{E. Holmgren}
\affiliation{Nanostructure Physics, Royal Institute of Technology, Stockholm, Sweden}
\author{A. Bondarenko}
\affiliation{Nanostructure Physics, Royal Institute of Technology, Stockholm, Sweden}
\affiliation{Institute of Magnetism, National Academy of Sciences of Ukraine, Kyiv, Ukraine}
\author{B.A. Ivanov}
\affiliation{Institute of Magnetism, National Academy of Sciences of Ukraine, Kyiv, Ukraine}
\affiliation{National University of Science and Technology MISiS, Moscow, 119049, Russian Federation}
\author{V. Korenivski}
\affiliation{Nanostructure Physics, Royal Institute of Technology, Stockholm, Sweden}

\date{\today}

\begin{abstract}
Vortex pairs in magnetic nanopillars with strongly coupled cores and pinning of one of the cores by a morphological defect, are used to perform resonant pinning spectroscopy, in which a microwave excitation applied to the nanopillar produces pinning or depinning of the cores only when the excitation is in resonance with the rotational or gyrational eigenmodes of the specific initial state of the core-core pair. The shift in the eigenmode frequencies between the pinned and depinned states is determined experimentally and explained theoretically, and illustrates the potential for multi-core spin-vortex memory with resonant writing of information on to various stable vortex pair states. Further, it is shown how the same resonant spectroscopy techniques applied to a vortex pair can be used as a sensitive nanoscale probe for characterizing morphological defects in magnetic films.
\end{abstract}

\maketitle

\section{\label{sec:intro}Introduction}

Spin vortices in ferromagnetic films are characterized by two topological charges. The first is the direction of the in-plane circulation of the spins in the vortex periphery about the vortex core, the so-called vortex chirality, which can be clock- or counterclock-wise. The second topological charge is the so-called vortex polarity, determined by the orientation of the out-of-plane spins comprising the core, which can be either "up" or "down".\cite{Wachowiak2002,Shinjo2000} The core itself is point-like, about 10 nm in size in Permalloy, but its motion effectively represents the nature of the dynamics of the vortex state as a whole.\cite{Choe2004,Vansteenkiste2009,Noske2016,Petit-Watelot2012} Core motion can be resonantly excited using either high-frequency magnetic field\cite{Bohlens2008,Waeyenberge2006,Ivanov2010,Buess2004} or spin-polarized current through the ferromagnet.\cite{Thomas2007,Moriya2008,Bolte2008,Kasai2006} The energetically tight and highly localized magnetic configuration of a spin vortex with its unique spin dynamics has been shown to yield high-quality-factor oscillators.\cite{Pribiag2007} Efficient control by dc, rf, and pulsed fields, as well as by spin-torque-currents, together	 with good thermal stability makes spin vortices a candidate system for non-volatile memory applications. Here, proposed data writing schemes include resonantly reversing the polarity of the vortex core\cite{Pigeau2010} and pinning vortices using antiferromagnetic layers with subsequent thermal control of the state using the N{\`e}el transition in the antiferromagnet.\cite{Araujo2016} A related memory idea based on resonantly depinning vortex-like domain-walls in a nanowire, the so-called race-track memory, was demonstrated to efficiently switch under spin-torque current excitation.\cite{Parkin2008}

A number of recent studies focused on lateral exchange-coupled vortex pairs\cite{hata1} and magnetostatically-coupled vortex arrays\cite{awad1, Galkin2006,Awad2010,Sugimoto2011,Shibata2004} as well as on vertical vortex pairs\cite{lebrun1,Stebliy2017,Haenze2016} have shown that the inter-vortex coupling can significantly alter the vortex-pair dynamics, giving rise to new resonant modes of motion, not present in the isolated vortex case. We have previously shown that in the limit of strong core-core coupling in a vertical stack, with a monopole-like interaction of the two cores, the collective resonance mode is an anti-phase rotation of the two cores about their 'center-of-mass' at a frequency that is increased ten-fold compared to the respective single-core gyrational modes in the system.\cite{cherepov1}

\begin{figure}[!t]
\includegraphics[width=3.4in]{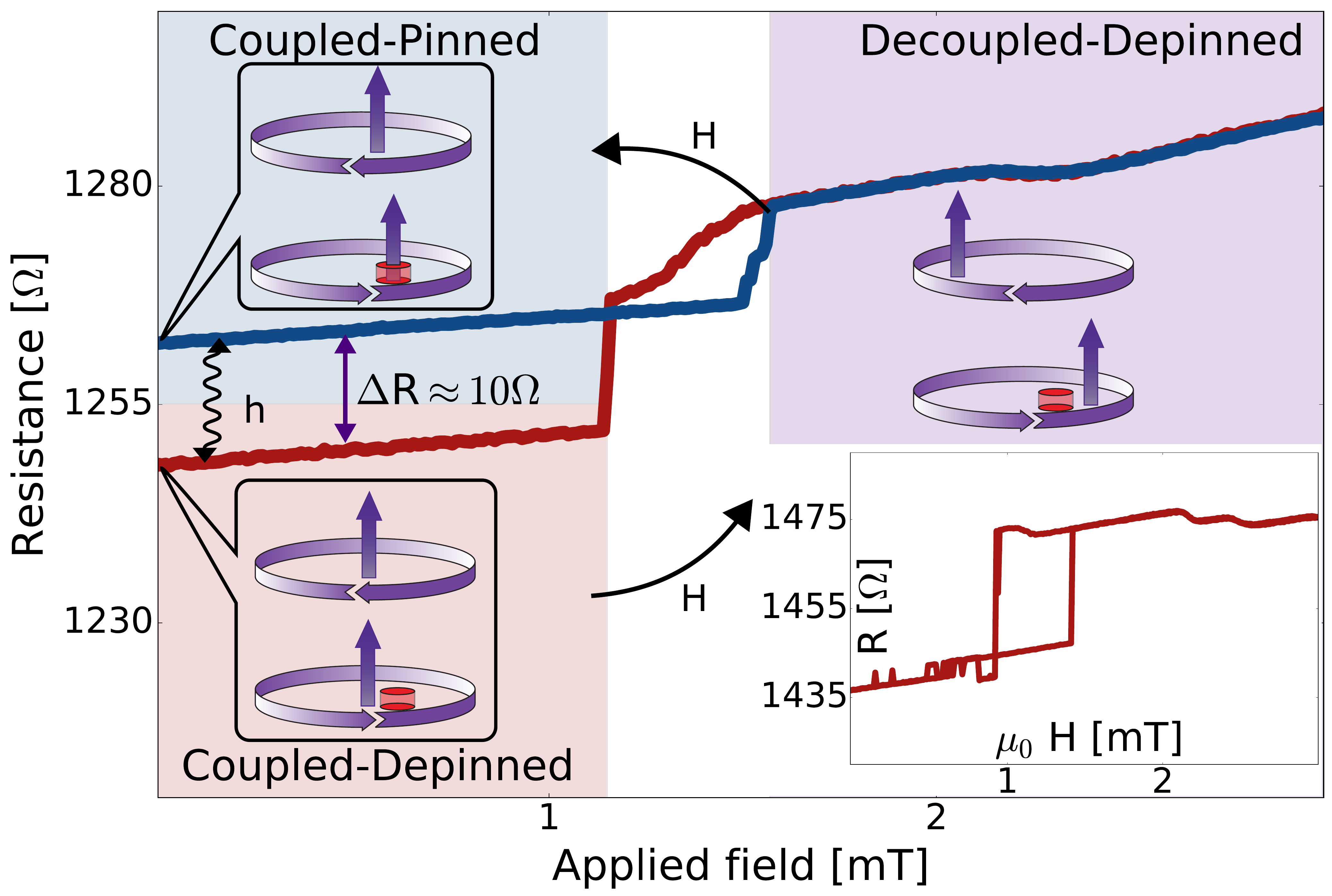}
\caption{Resistance versus in-plane field for a SAF magnetic junction in the parallel-core antiparallel-chirality (P-AP) vortex pair state. Zero-field hysteresis (of about 10 $\Omega$, with R reflecting the position of the vortex core) is induced by a pinning site (depicted as a red dot) located near the center of the particle and is superposed onto the core-core coupling hysteresis, which is shown in the inset for an "ideal" sample with no pinning and no hyseresis at zero field (a different junction from the same fabrication series). The spin configuration of the vortex pair is schematically shown at different points in the sweep: the circular arrows indicate the direction of the vortex chirality, while the vertical arrows indicate the cores' polarization and location. As the dc-field (H) is increased above 1.7 mT, the two vortex cores decouple (illustrated above the inset with the purple-shaded background). At low dc-fields (below 1 mT) the cores are strongly coupled, with one of the cores either pinned (blue shade) or depinned (red shade), with the system switching between these two stable states (between the red and blue R-branches at H=0) under resonant ac-field (h).}
\label{fig1}
\end{figure}

Pinning of isolated spin vortex cores, both on material-intrinsic low-pinning-energy imperfections and on nanofabricated high-pinning-energy defects,\cite{Burgess2013} has been investigated. It was shown that the dynamics of a pinned isolated vortex can change dramatically compared to the unpinned case.\cite{Compton2006,compton2010,Chen2012,Chen2012a} Core-pinning of vortex pairs, however, remains largely unexplored.

In this work we show how a vortex pair of parallel core polarization and antiparallel chirality (P-AP) can be pinned to and depinned from an intrinsic defect by resonantly exciting the dynamics of the vortex pair by an external ac-field, which yields multiple stable, memory-viable states with reliable interstate switching. The pinned versus depinned state is read out by measuring the zero-field hysteresis in a suitably designed magnetic junction, as shown in Fig.\ref{fig1}. The zero-field hysteresis due to pinning is superposed onto the core-core decoupling/recoupling hysteresis shown in the inset to Fig.\ref{fig1}, between 1 and 1.5 mT for this specific junction, where the strong core-core attraction is overcome by the effect of the static field (H). The pinning is found to modify the resonance frequencies of the core-core motion compared to the depinned case, which allows unidirectional switching and makes the system a candidate for vortex-based memory. Furthermore, we theoretically explain the resonance spectra and the observed switching thresholds, and show how the spectrum of the system can be used to sensitively probe the location, type, and energetics of defects. This allows the vortex pair to be used as a nanoscale probe for material defects and device imperfections. 

\subsection{\label{sec:samples}Samples and Measurements}
The free layer of the junctions measured consists of two elliptical particles, a so-called synthetic antiferromagnetic (SAF) pair, of 5 nm thick permalloy (Ni$_{80}$Fe$_{20}$), separated by a tantalum-nitride spacer with a thickness of 1 nm, which fully suppresses the exchange coupling between the layers. The junctions were designed with lateral dimensions ranging from 350 to 420 nm with an aspect ratio of 1.2. The state of the SAF is read out resistively via a tunnel barrier separating it from an exchanged biased reference SAF. The small stray field from the reference layer and a thickness imbalance, if any, between the two permalloy layers\cite{Koop2014} were determined to be negligible in the present study of vortex core pinning effects. The samples were fabricated and measured using methods described elsewhere.\cite{samples2,samples1}

GHz range ac-field excitations of given amplitude and frequency were applied in bursts of 300 ms using on-chip integrated waveguides. The resistance of the junctions and its field-hysteresis were used to determine the specific states of the vortex pair. Prior to the measurements, the two states at zero field were characterized using microwave spectroscopy to determine which one is pinned/depinned. In this, low ac field amplitudes were used to avoid switching (de/pinning) and the spectra were compared to those expected theoretically as well as those measured on samples without zero-field core pinning. All data were collected at 77 K and zero bias field, where the studied vortex-core hysteretic properties dominate the behavior of the system.


\begin{figure}[!t]
\includegraphics[width=3.4in]{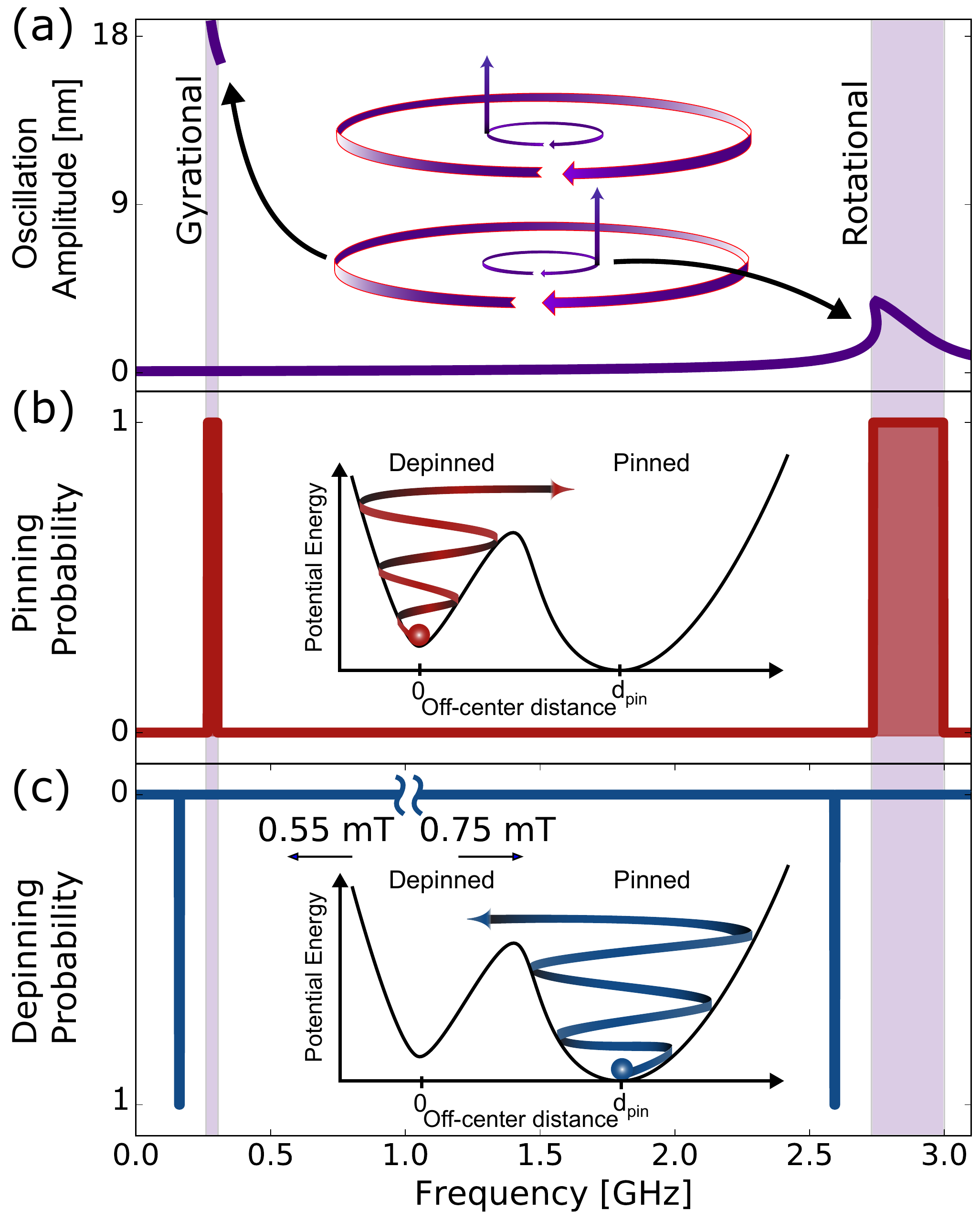}
\caption{(a) Theoretical forced oscillation spectra of a vortex pair, with two characteristic resonance modes at about 300 MHz, large-radius gyrational, and 3 GHz, small-radius rotational; the inset shows the corresponding trajectories with the circular arrows, and the two cores rotating in anti-phase with the vertical arrows. Measured microwave spectra for the probability of pinning (b) and depinning (c) of the core-core pair for a SAF junction in the P-AP vortex state. In the case of depinning (c) the field-amplitude above 1 GHz was increased to show a clear response at the rotational eigenmode. When the vortex pair is not pinned, the resonance frequencies correspond to the gyrational and rotational eigenmodes of the system (red bars; overlap with theory -- purple bands). When the pair is pinned the magnetic structure of the pinned core is changed, which shifts the resonance frequencies (blue bars).}
\label{fig2}
\end{figure}

\section{Theory}

The dynamics of the P-AP vortex state is well described by the Thiele equations\cite{Thiele1972} for the planar motion of the individual vortex cores subject to an interaction potential, $U(\mathbf{x}, \mathbf{X})$, with the resulting Lagrangian of the form
\begin{eqnarray}
\mathcal{L} = \frac{G}{2} x \dot{y} + 2 G X \dot{Y} - U(\mathbf{x}, \mathbf{X}).
\end{eqnarray}
Here, the layers are assumed symmetric, $\mathbf{x=X_1-X_2}$ is the core-core separation of the pair, $\mathbf{X=(X_1+X_2)/2}$ -- the pair's offset from the center of the particle, $\mathbf{X}_i$ -- the position of the core in layer i, and G -- the gyroconstant. 

The potential energy of the P-AP state is the sum of the core-core attraction, the magnetostatic repulsion from the particle boundary, and the Zeeman energy:
\begin{eqnarray}
\begin{aligned}
U(\mathbf{X}_1,\mathbf{X}_2) &= U_{c-c}(\left|\mathbf{X}_1 - \mathbf{X}_2\right|) + \frac{k_1}{2} \mathbf{X}_1^2 + \frac{k_2}{2} \mathbf{X}_2^2 +\notag\\
& -\frac{\pi}{2}\mu_0 M_s L_z \left[ \varepsilon_{jk}(\mathbf{X}_1)_j H_k - \varepsilon_{jk}(\mathbf{X}_2)_j H_k\right].
\end{aligned}
\end{eqnarray}
The core-core coupling energy has been discussed in detail in our earlier study\cite{cherepov1} and is the sum of the pairwise interactions of the four individual surface poles of the two vortex cores. This monopole-like form, as against dipole-dipole coupling, is needed to accurately represent the near-field interaction through the ultra-thin spacer in our system with strong direct core-core coupling. The repulsion of the core from the particle boundary is characterized by the stiffness constants given as\cite{Guslienko2002} $k_i=20\mu_0M_{s,i}^2L_{z,i}^2/9L$, where $L_z$ is the length of the core and $L$ -- the lateral size of the particle. The strong core-core attraction and the core-boundary repulsion result, \emph{specifically for AP-chirality vortex pairs, in contrast to P-chirality pairs or single vortices}, in a deep potential minimum in the center of the particle. The Zeeman energy term is of opposite sign in the two vortices due to their antiparallel chiralities, and $\varepsilon_{jk}$ is the two-dimensional Levi-Cevita symbol.

The forced oscillation of a core-pair with zero center-of-pair offset, $\mathbf{X}=0$, under an ac-excitation of frequency f and normalized amplitude $\hat{h}=2Lh/(4\pi M_s \Delta)$, with $\Delta \approx 8$ nm -- the core radius in permalloy, is described by
\begin{eqnarray}
f=\frac{\omega(a)}{1+\lambda^2}\left[ 1\pm \sqrt{1-(1+\lambda^2)\left( 1-\frac{\hat{h}^2}{\omega^2(a)a^2} \right)} ~ \right].
\label{papspectrum}
\end{eqnarray}
Here $\omega(a)=(\partial U_{c-c}(a) / \partial a)/2G a$ is the intrinsic frequency of the core gyration (normalized by $\gamma M_s$), with the core-core separation $a$, and $\lambda$ is the dissipation constant. The offset can be taken as constant since the symmetry of the system allows only weak to negligible parametric excitation of the center-of-pair motion. 

As the excitation amplitude is increased, the dynamic response at the high-frequency rotational mode, $\omega_r \approx $ 2-3 GHz for our samples, extends to lower frequencies, while the low-frequency gyrational mode, $\omega_g \approx $ 200-300 MHz, extends to higher frequencies, as shown in Fig.\ref{fig2}(a) (with more detail in section \ref{sec:results} below and Fig.\ref{fig5}(b) inset). This takes place due to the fact that the intrinsic frequency of the system decreases as the radius of the core trajectory increases. 

Pinning of a P-AP vortex-pair can occur on either intrinsic defects or surface roughness. Intrinsic defects, such as various grain boundaries\cite{OHandley1999} in polycrystalline films, lower the local exchange stiffness, $A$, which reduces the energy penalty of having the out-of-plane directed core at this location. At the same time, such defects typically result in a lower magnetization, which reduces the stray fields from the vortex core. 

Surface roughness in the form of a local variation in the magnetic film thickness reduces the exchange energy for the out-of-plane spins as well the stray field from the core and the core-core interaction caused by the change in the core length. Such thickness variations can occur at either interface of each magnetic layer, towards or away from the other magnetic layer. The specific changes in the interaction and stray field energies thus depend on exactly which interface is affected and how. 

For the above defects to be effective in pinning a vortex core, their lateral size must be comparable to the size of the core. The model below therefore assumes, as the first approximation, that a given pinning site is comparable or somewhat larger than the vortex core size, $L_{pin} \gtrsim \Delta$.

Pinning of one core to a defect displaced from the center of the particle leads, via the strong core-core coupling, to an effective pinning of the second vortex core, such that the core-core pair as a whole is displaced off center, as long as the pair remains in the coupled state. The core-core separation, being zero in the equilibrium unpinned state, becomes finite in the pinned state of the pair. As a result, the dynamics of the pair changes, also affected by the modification of the pinned core's spin distribution outlined earlier. Assuming a linear dynamic regime, with the core motion being a small-amplitude oscillation near the bottom of a specific pinning potential, the linearized equations of motion take the form: 
\begin{eqnarray}
\begin{aligned}
G_{pin} \mathbf{e}_z \times \dot{\mathbf{x}}_1 &= - \kappa_{pin} (\mathbf{x}_1-\mathbf{x}_2) - k_{1,pin} \mathbf{x}_1\\
G \mathbf{e}_z \times \dot{\mathbf{x}}_2 &= - \kappa_{pin} (\mathbf{x}_2-\mathbf{x}_1) - k_2 \mathbf{x}_2.
\end{aligned}
\end{eqnarray}
The coefficients depend on the characteristics of the pinning site affecting the spin structure of the pinned core. The gyrovector is given by
\begin{eqnarray}
\mathbf{G}=\int dV \left[ \frac{M_s}{\gamma} \left( \nabla \phi \times \nabla \theta \right) \sin \theta \right],
\label{gyrofactor}
\end{eqnarray}
where $\phi$ and $\theta$ are the polar and azimuthal angles of the magnetization. It is dependent on both the local magnetic film thickness and the local saturation magnetization of the pinned vortex core, changing in magnitude when the core is captured by the pinning site. Assuming the core magnetization is axially symmetric the gyroconstant can be rewritten as
\begin{eqnarray}
\mathbf{G}=\mathbf{e}_z \frac{2 \pi}{\gamma} \int d(M_{s,pin}\cos \theta ) L_{z,pin}
\end{eqnarray}
in cylindrical coordinates. Assuming constant saturation magnetization throughout the core, the gyroconstant scales as
\begin{eqnarray}
G_{M_{s,pin}} = G_0 \frac{M_{s,pin}}{M_s}.
\end{eqnarray}
Here $G_0=\mu_0 L_zM_s/2\gamma$ is the unperturbed gyroconstant. Should the saturation magnetization not be homogeneous through-out the core, the first order correction would take the form of a constant form-factor dependent on the relative size of the core versus the pinning site, which would not change the predicted behavior except for a quantitative correction. Similarly, the change in the gyroconstant for pinning by roughness can be written as
\begin{eqnarray}
G_{rgh} = G_0 \left( 1+  \frac{\gamma + \delta}{L_z} \right),
\end{eqnarray}
with the surface pinning defect height at the interface toward the other magnetic layer $\gamma$ and away from it $\delta$.

\begin{figure}[!t]
\includegraphics[width=3.4in]{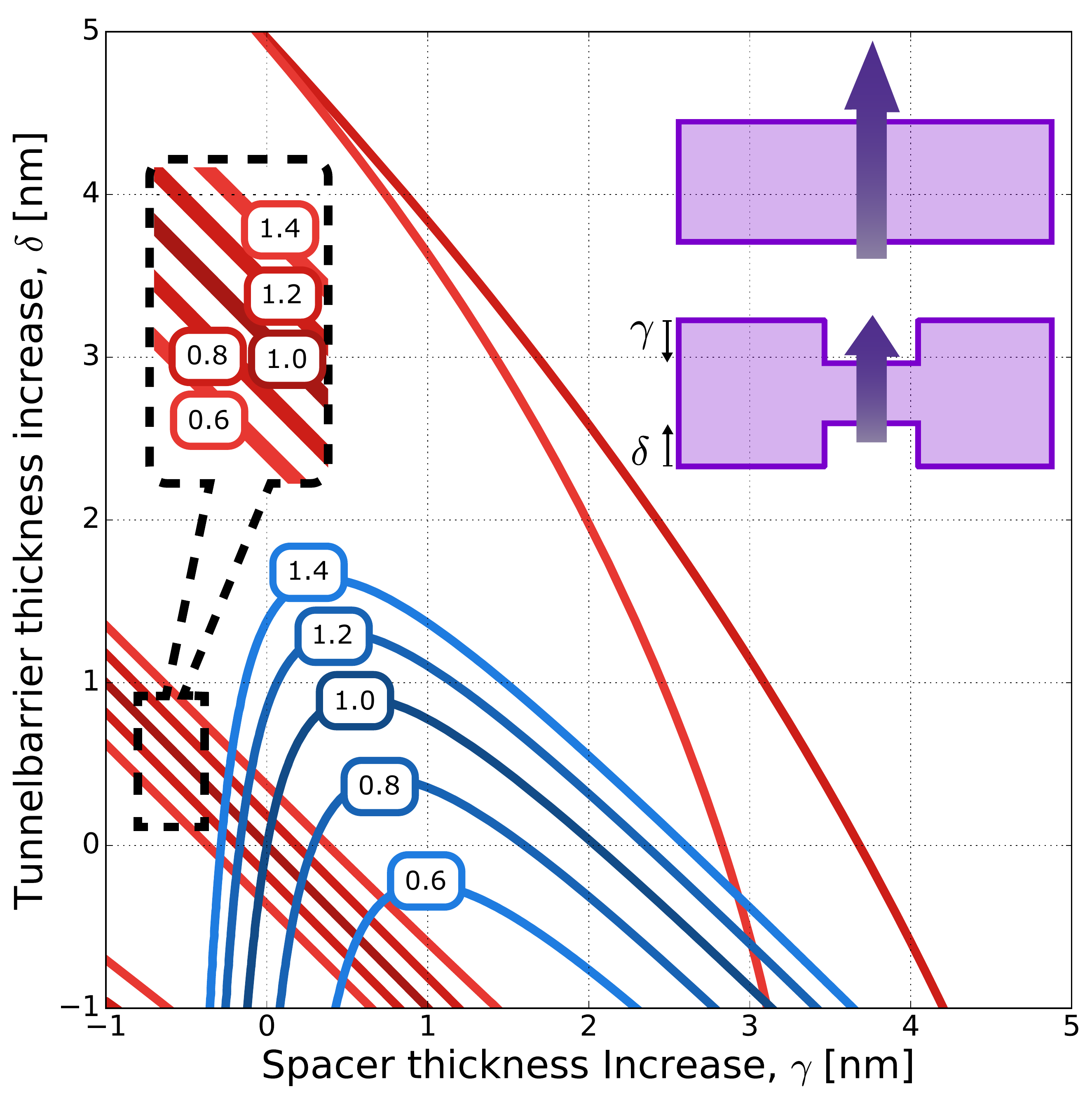}
\caption{Theoretical contour lines of the relative change in frequency, $f_{i,pin}/f_i$, of the gyrational (red curves) and rotational (blue curves) resonance modes as a function of the roughness parameters at the two interfaces of the bottom layer ($\gamma$ at top and $\delta$ at bottom, as illustrated in the inset). If the total thickness of the layer is reduced, the corresponding reduction of the gyroconstant increases the frequency of the gyrational resonance. Should either pole of the pinned core be moved away from the other core, the total core-core coupling is weakened, which reduces the frequency of the rotational resonance. The limits of the axis were chosen at $\gamma$ and $\delta$ equal to -1 and +5 to avoid the unphysical regions where the spacer/tunnel-barrier (at -1 and below) or the vortex layer (at 5 and above) would completely vanish. The red contours close outside the region shown (below -1 in $\delta$).}
\label{fig3}
\end{figure}

The linearized core-core interaction is characterized by the stiffness constant:
\begin{eqnarray}
\begin{aligned}
\kappa = \frac{\mu_0 M_s^2 \Delta^2}{2} &\left( \frac{1}{D} + \frac{1}{L_{z,1}+L_{z,2}+D}  \right. \\ &\left.  -\frac{1}{L_{z,1}+D} - \frac{1}{L_{z,2}+D} \right),
\end{aligned}
\end{eqnarray}
where $D$ is the lateral separation of the two vortex cores.

The interaction with the particle boundary is not changed, but there is an additional restoring force due to the pinning-induced variation in the exchange energy of the core. This is taken into account by modifying the stiffness constant:
\begin{eqnarray}
k_{rgh} = k_0 \left( 1+\alpha \frac{\gamma + \delta}{L_z} \right),
\end{eqnarray}
where coefficient $\alpha$ is proportional to the ratio of the stiffness of the core interaction with the pinning site and the particle boundary. As shown below, based on our decay measurements of the pinning potential and the broadening of the rotational resonance, the pinning energy of the site measured in Fig.\ref{fig2} is approximately 170 meV and is located about 10 nm off the particle center. Assuming a linear pinning-potential versus distance dependence and the pinning-site size equal to the core size, the pinning stiffness can be estimated to be $k_{rgh}\approx 1.5\cdot 10^{-4}$ N/m. The sample geometry dictates the boundary stiffness\cite{cherepov1} of $k_0\approx 1.1 \cdot 10^{-4}$ N/m, which results in $\alpha \approx 1.4$. 

The eigenfrequencies of the system can be calculated using the exponential ansatz, $\mathbf{X}_i (t)=\mathbf{X}_{i,0} \exp (i\omega t)$, and solving the eigenvalue problem for $\omega$:
\begin{eqnarray}
\begin{aligned}
\omega^2 = \frac{1}{2}\left[\Omega_1^2 + 2 \omega_1 \omega_2 + \Omega_2^2\right. \\
\left.\pm(\Omega_1+\Omega_2)\sqrt{(\Omega_1-\Omega_2)^2+4\omega_1 \omega_2}\right],
\end{aligned}
\end{eqnarray}
where $\omega_i=\kappa/G_i$ and $\Omega_i=\omega_i + k_i/G_i$. The gyrational resonance is given as $\omega_g=k/G$, while the rotational as $\omega_r=(2\kappa+k)/G$. The resonant frequencies versus a small change in the saturation magnetization, $\varepsilon_{M_s} = 1-M_{s,pin}/M_s$, are then given by
\begin{eqnarray}
\begin{aligned}
\omega_{g,pin} &=\omega_g \sqrt{1+\varepsilon_{M_s}} \\
\omega_{r,pin} &=\omega_r \sqrt{1-\varepsilon_{M_s}\frac{\omega_r-2\omega_g}{\omega_r}}.
\end{aligned}
\label{msshift}
\end{eqnarray}
As a result, for pinning by a defect with reduced local magnetization, the gyrational resonance frequency shifts up due to a lower gyroconstant, while the rotational frequency shifts down due to a weaker stray field from the pinned core and hence a weaker core-core coupling.

\begin{figure*}[!ht]
\includegraphics[width=7in]{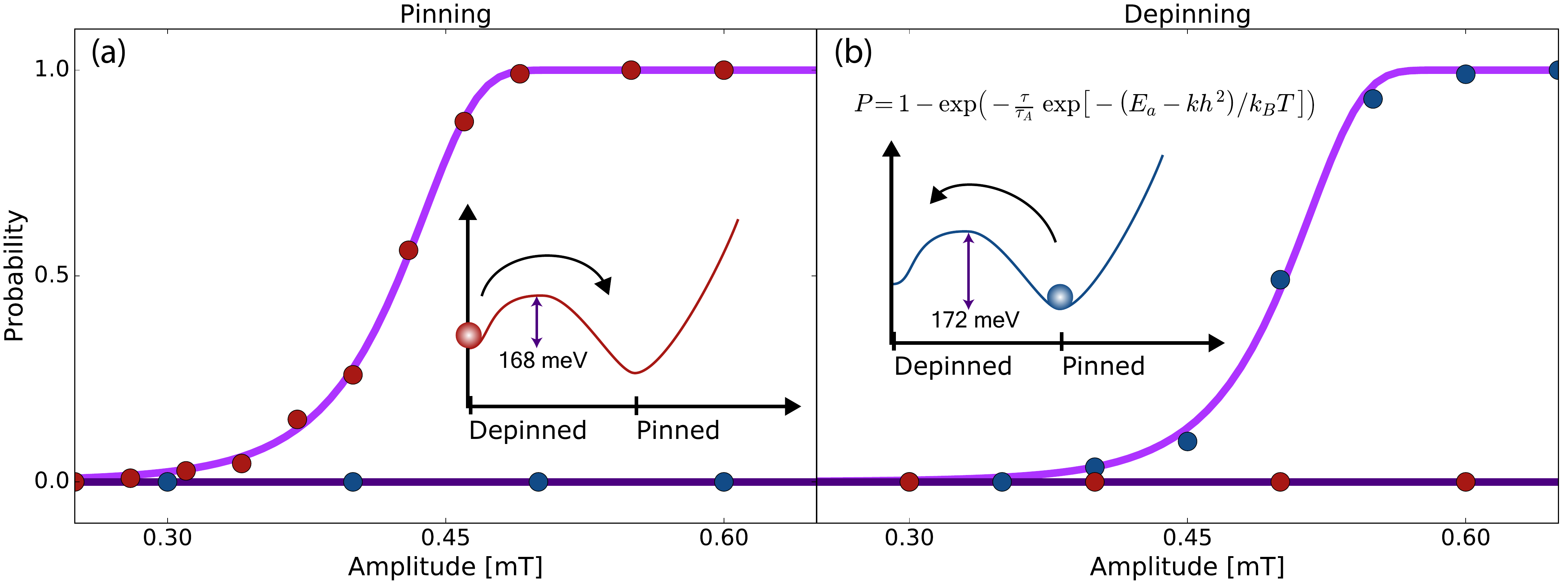}
\caption{Measured probability of resonantly excited transition from depinned to pinned (a) and pinned to depinned (b) vortex-pair states versus the excitation amplitude at 160 MHz (blue circles) and 285 MHz (red circles). The individual data point error bars are smaller than the symbol size. The purple line is the fit to the data using Eq.(\ref{eqfit}). The extracted activation energy values in the two cases are shown in the insets to (a) and (b) depicting, respectively, the coupled-depinned state with the cores centered in the particle (corresponding to the red-shaded area in Fig.\ref{fig1}), and the coupled-pinned state slightly off the particle center (corresponding to the blue-shaded area in Fig.\ref{fig1}). The pinning energy is given by the potential depth in the "depinning" case.}
\label{fig4}
\end{figure*}

The spectral changes caused by pinning to surface roughness depend on the relative thickness change and which interface is more affected. Qualitatively, the effect of a vortex core pinning by roughness on the gyrational frequency is straight forward -- the gyroconstant is proportional to the thickness of the layer, so for roughness where the effective thickness is increased the frequency decreases, and vice-versa. The effect of roughness on the rotational resonance depends more on precisely which interface is affected and how specifically the morphology of the interface is changed. If the thickness is perturbed at the interface facing the other vortex, such that the separation of the cores is increased (thicker spacer), the rotational frequency is decreased due to a weaker core-core interaction. In the opposite case, for a thinner spacer, the rotational frequency is increased. If the thickness is perturbed at the opposite interface, so that the far end of the pinned core extends away from the other vortex core (roughness of the bottom interface extending into the tunnel barrier), the total core-core interaction is weaker, which decreases the rotational frequency. Naturally, for the opposite case of the inward roughness of the bottom interface, the rotational frequency is higher. The quantitative changes are shown in Fig.\ref{fig3} for the typical case of $\alpha=1$. The results using the estimated value of $\alpha \approx 1.4$ for our system are within 10\% of those shown in Fig.\ref{fig3}.

We find experimentally that this frequency shift is sufficient to selectively and reliably change the vortex pair state from pinned to depinned and, vice-versa, from depinned to pinned, by applying resonant excitation pulses at the corresponding frequencies.

In order to model this process we note that an earlier, conceptually relevant study of thermal escape from the zero voltage state of a Josephson-junction has shown that the effective activation energy is reduced under resonant excitation.\cite{Devoret1987} Similarly, the effective activation energy of a pinning-site is reduced by the resonant excitation of the vortex-pair to an oscillation trajectory within the pinning site, with the steady-state energy above the ground state. At low amplitudes, it can be shown that the radius of such a trajectory scales linearly with the ac-field amplitude and the corresponding reduction in the activation energy scales as the square of the radius. The unperturbed activation energy can be determined by fitting the probability of resonant pinning and depinning versus the excitation amplitude to
\begin{eqnarray}
P= 1-\exp \left( -\frac{\tau}{\tau_A} \exp \left[ -\frac{1}{k_BT} \left(E_a-kh^2 \right) \right] \right),
\label{eqfit}
\end{eqnarray}
which is a Poisson probability, taking into account the modified activation energy. Under resonant excitation, thermal fluctuations pin/depin the pair at some attempt rate, $\tau_A$, during the excitation duration, $\tau=300$ ms. When the excitation is in resonance, the attempt rate is equal to the inverse of the initial states' gyrational frequency.

\section{\label{sec:results}Results and discussion}

As the coupled cores of a P-AP vortex pair in our system are made to move within the respective magnetic layers, either core can be pinned to a defect, which in turn pins the other core and the pair as a whole due to the strong core-core coupling. Large-amplitude core motion only occurs when the excitation is in resonance with either the gyrational or rotational resonant modes of the pair, as shown in Fig.\ref{fig2}(a). This explains why pinning of the pair from an initially depinned state is experimentally observed at these resonant frequencies,  $\sim$300 MHz and $\sim$3 GHz, shown in Fig.\ref{fig2}(b), and not observed at off-resonance frequencies.

\begin{figure*}[!t]
\includegraphics[width=7in]{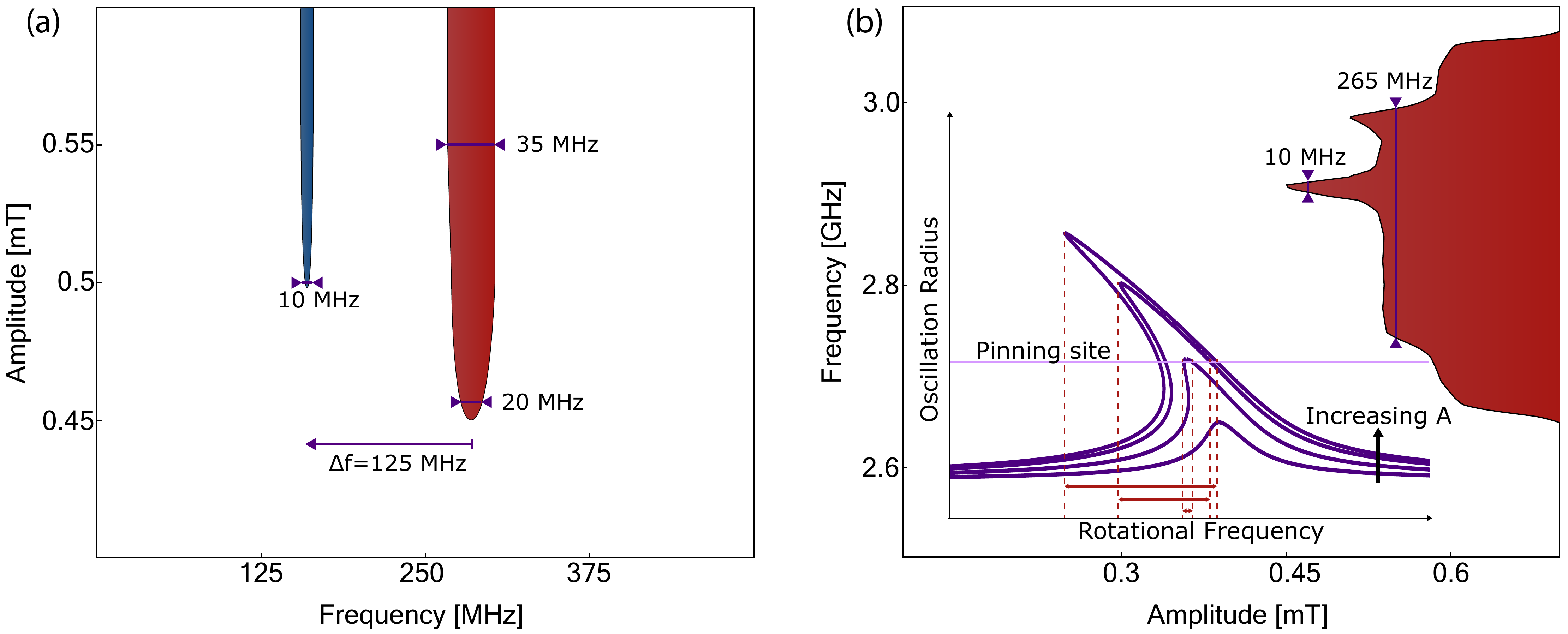}
\caption{(a) Pinning (red) and depinning (blue) switching maps in the vicinity of the gyrational resonance of a P-AP vortex pair. The measured frequency shift, together with Eq.(\ref{msshift}) and Fig.\ref{fig3}, can be used to determine the nature of the pinning site. The frequency shifts in this case are consistent with roughness-induced pinning, where the spacer thickness is increased by 0.1 nm and the tunnel barrier thickness is decreased by 0.4 nm, within an area roughly of the size of the vortex core. (b) Pinning map at the rotational resonance as a function of the ac-field amplitude. The inset shows the theoretically expected broadening and tilting of the rotational peak as the amplitude is increased. The threshold at which the resonance starts to rapidly broaden can be used to determine the trajectory radius corresponding to the location of the pinning site with respect to the particle center.}
\label{fig5}
\end{figure*}

As discussed above, the dynamics of the pinned pair are modified due to changes in the pinned core's magnetic properties. Pinning detunes the vortex pair from the original core-core resonance frequency such that an excitation at a different frequency is required to depin the pair. This is indeed what is observed on the experiment for our samples, with typical frequency shifts of 100 to 200 MHz at the gyrational resonance and up to 500 MHz at the rotational resonance, as shown in Fig.\ref{fig2}(c). The insets to Fig.\ref{fig2}(b,c) show the potential landscape as the distance from the particle center is increased, with the pinning site located at distance $\textnormal{d}_{\textnormal{pin}}$. We note that, for strongly core-core coupled vortex pairs with AP-chiralities, the depinned pair has a deep potential minimum at the center of the particle, which acts as a second stable state for the probability measurements.

Rather detailed characteristics of a pinning center can be inferred from studying the resonance spectra of a core-core coupled vortex pair. Fig.\ref{fig4} shows the pinning and depinning probability data as a function of the ac-field amplitude, with the ac-field frequency corresponding to the respective gyrational resonance. The data fitted using Eq.(\ref{eqfit}) yield the pinning potential to be about 170 meV, which is generally consistent with the previous studies of vortex pinning to intrinsic material defects.\cite{compton2010,Burgess2013,Chen2012a}  

Vortices are pinned by a reduction in the effective energy of the core for two main types of material defects: either an intrinsic imperfection, locally reducing the saturation magnetization; or film roughness, locally changing the thickness of the particle. The type of a specific defect can be inferred from comparing the measured shift in the resonant frequencies between the pinning and depinning of the core-pair, such as shown in Fig.\ref{fig5}(a) for the gyrational resonance and Fig.\ref{fig2} for the rotational resonance, to the behavior expected for that a type of a pinning site, described by Eq.(\ref{msshift}) and Fig.\ref{fig3}. 

To reiterate, if the pinning is due to a reduced magnetic moment of the pinned core, the core-core coupling is weakened and the rotational resonance is shifted to lower frequencies while the gyrational frequency is increased due to a decrease of the gyroconstant. 

Pinning due to roughness can, in all but very select few cases, be distinguished from pinning due to a reduction of the core's magnetic moment. The details of how roughness-type defects affect the pinned-core dynamics depend sensitively on the defect morphology, as discussed in the theory section above. In many cases, the specific effects on the measured spectra and the depinning/pinning probability are unique. The only exception is a roughness-type defect, for which the local effective layer thickness is decreased and the core-core distance is increased. In this case the effect on the pinned-core-pair spectrum should be similar to that due to an intrinsic imperfection with reduced magnetization and a more careful analysis is required to distinguish the two types of pinning sites. 

As an illustration, the data in Fig.\ref{fig5}(a) show that the gyrational frequency is shifted down by 125 MHz, $f_{g,pin}/f_g\approx 0.6$, for pinning the vortex-pair and the rotational frequency (Fig.\ref{fig2}) is shifted down by 350 MHz, $f_{r,pin}/f_r\approx 0.9$. Such spectral modification can only occur if the core is pinned by a roughness defect, which extends the pinned core away from the other core by an average of 0.4 nm at the far-interface ($\delta = -0.4$) and 0.1 nm at the near-interface ($\gamma = 0.1$), with the lateral size of the thickness perturbations comparable to the size of the vortex core. The defect can then be identified as a nano-sized "dent" in the tunnel barrier, which is smoothed out by the permalloy layer (potentially non-wetting and fine-grained\cite{samples1}) and further smoothed out by the tantalum nitride spacer, such that the "dent" is too small to pin the second vortex core in the top permalloy layer.

The significant difference in the bandwidth of the gyrational resonance between the pinned and depinned states of the vortex pair seen in Fig.\ref{fig5}(a) can be explained by the fact that the depinned vortex has more freedom of movement and can thus sustain a variety of low-energy gyrational trajectories. When the vortex becomes pinned, a large part of these low energy oscillations are locked out, which substantially reduces the bandwidth at low amplitudes for the pinned ground state. It is interesting to note that this behavior is opposite to that of the single vortex, where one would expect pinning sites to broaden the resonance bandwidth as several such sites are averaged over in a large-radius trajectory. The bandwidth narrowing in the vortex-pair case is due to the small radius of the rotational core motion in the coupled state, combined with a negligible 'center-of-mass' motion of the pair as a whole, together resulting in no pinning-site averaging effect (over, typically, multiple defects in a nanostructured material). The bandwidth reduction upon pinning could be an attractive property for nano-oscillators, where potentially inter- and intra-mode tuning (frequency and quality factor) could be realized using suitable pinning sites.

The location of the defect can be determined by the bandwidth of the resonant pinning map at the rotational resonance, shown in Fig.\ref{fig5}(b). The pair is not pinned until the core oscillation, at a given excitation amplitude, begins to overlap with the pinning site, which results in a threshold-like broadening of the resonant pinning probability at that amplitude. This threshold can be used to determine, through Eq.(\ref{papspectrum}), the radius of the trajectory and thus the pinning site's distance from the center of the particle. Due to the extremely small radius of the core oscillation in the rotational resonance, of the order of a few nanometer, the pinning site's location can be accurately determined. In the case of Fig.\ref{fig5}(b) the radius can be calculated to be about 8-10 nm.

The above analysis is a most comprehensive characterization of a typical pinning site, made possible by the unique topological and energetic structure of vortex pairs in our SAF structures: the defect is of vortex core size, located about 9 nm from the center of the particle, where the local thickness is changed by roughly 0.4 nm at the tunnel barrier interface and 0.1 at the spacer interface, both in the direction toward the tunnel barrier and away from the other vortex core, resulting in the core-pinning potential of 172 meV.

\section{\label{sec:conc}Conclusions}
In summary, we have experimentally examined the resonant properties of a spin vortex pair with parallel cores and antiparallel chiralities, pinned by an intrinsic defect. The induced dynamics under resonant excitation can be used to pin and depin the pair. The individual properties of the vortices as well as dominant core-core attraction is modified in the pinned state, which changes the frequency of the resonant modes in the system and offers the possibility of unidirectionally setting in a specific stable state of the vortex pair. This in turn could make possible multi-level memory implementations based on suitably core-pinned spin-vortex pairs. The observed shift in the resonant frequencies together with the bandwidth-reduction upon pinning the vortex pair could have relevance for vortex-based nano-oscillators, utilizing suitably engineered pinning sites for tuning the frequency range and quality factor. Additionally, through comparing the pinned and depinned microwave spectra of a vortex pair with the behavior calculated from a linearized Thiele-equation model, the position, characteristic pinning energy, and nature of a specific defect can be determined, making system an ultra-sensitive nanoscale probe for studying defects in magnetic nanostructures.

\section*{Acknowledgment}
Support from the Swedish Research Council (VR Grant No. 2014-4548), from the National Academy of Sciences of Ukraine via project number 1/17-N, and from the Ministry of Education and Science of Russian Federation in the framework of Increase Competitiveness Program of NUST MISiS (No. К2-2017-005), implemented by a governmental decree dated 16th of March 2013, N 211, are gratefully acknowledged.

\bibliographystyle{unsrt}
\bibliography{refs}

\end{document}